\documentclass[12pt]{article}
\usepackage{graphicx}

\def\ii{\'{\char'20}}
\def\beq{\begin{equation}}
\def\eeq{\end{equation}}
\def\bea{\begin{eqnarray}}
\def\eea{\end{eqnarray}}

\def\lapprox{\hbox{\lower .8ex\hbox{$\,\buildrel < \over\sim\,$}}}
\def\gapprox{\hbox{\lower .8ex\hbox{$\,\buildrel > \over\sim\,$}}}

\begin{document}
\begin{center}
{\Large \bf The variation of the gravitational constant inferred 
            from the Hubble diagram of Type Ia supernovae}  \\
\vspace{4 mm}

E. Garc{\ii}a--Berro\dag\ddag, 
Y. Kubyshin\footnote{On leave of absence from the 
Institute  of  Nuclear Physics, Moscow State University \\
119992 Moscow, Russia. E-mail: iouri.koubychine@upc.edu}$\S$, 
P. Lor\'en--Aguilar\P\ddag\
\& J. Isern\P\ddag

\vspace{4 mm}

\dag\  Departament de F{\ii}sica Aplicada, Universitat Polit\`ecnica  
       de Catalunya,  \\
       Escola Polit\`ecnica Superior de Castelldefels,\\
       Avda. del Canal Ol\'\i mpic s/n, 08860 Castelldefels, Spain \\
\ddag\ Institut d'Estudis Espacials de Catalunya, \\
       Ed. Nexus, c/ Gran Capit\`a 2, 08034 Barcelona, Spain \\
$\S$\  Institut de T\`ecniques Energ\`etiques, Universitat Polit\`ecnica 
       de Catalunya, \\
       Edif. ETSEIB, Campus Sud, Avda. Diagonal, 647, 08028 Barcelona, Spain \\
\P\    Institut de Ci\`encies de l'Espai, C.S.I.C., \\
       Campus UAB, Facultat de Ci\`encies, Torre C--5,  
       08193 Bellaterra, Spain\\
\end{center}

\begin{abstract}
We  consider  a  cosmological  model  with  a  variable  gravitational
constant,  $G$, based  on a  scalar-tensor theory.   Using  the recent
observational  data  for the  Hubble  diagram  of  type Ia  supernovae
(SNeIa) we find a phenomenological expression describing the variation
of $G$.   The corresponding variation  of the fine  structure constant
$\alpha$  within multidimensional  theories  is also  computed and  is
shown not to support known constraints on $\Delta \alpha / \alpha$.
\end{abstract}

\section{Introduction}

The analysis of the observations of the Hubble diagram of distant type
Ia supernovae  \cite{Riess98, SCP, Riess04} provide  evidence that the
universe has been accelerating  recently, at $z<0.5$, and decelerating
at earlier  stages \cite{Riess04, Turner02}.   The Friedmann cosmology
without cosmological constant and with zero curvature --- as indicated
by the recent CMB  Boomerang and Maxima experiments \cite{Boo, Maxima}
--- cannot explain  such an evolution of  the universe \cite{Ratra03}.
The accelerated behaviour can be  attributed to a ``dark energy'' with
negative pressure, the simplest  possibility being the introduction of
the cosmological constant in  accordance with cosmic concordance model
$\Omega_{M}  \approx  0.3$,  $\Omega_{\Lambda}  \approx 0.7$  ---  see
\cite{Sahni00} for a review.

An  alternative solution is  to modify  the gravitational  theory, for
example,  by allowing  a  time variation  of  Newton's constant.   The
possibility of a time variation of fundamental constants of nature, in
particular  of  the fine  structure  constant,  $\alpha$,  and of  the
gravitational  constant, $G$  ---  first considered  by  Dirac in  the
framework  of his  Large  Number hypothesis  \cite{Dirac37} and  later
developed in  \cite{BD61} within an alternative  theory of gravitation
(see references \cite{Uzan03} and  \cite{Will93} for more details) ---
has  been recently  a subject  of  numerous studies  (see for  example
references \cite{Uzan03, Damour03,  GIKL04, Uzan04} for recent reviews
and  extensive  bibliography).   It  is  worth  mentioning  that  many
theoretical approaches,  such as models with  extra dimensions, string
theories or scalar--tensor models of quintessence, contain a built--in
mechanism   for  a   possible   time  variation   of  the   couplings.
Astronomical  measurements   allow  to  constrain   such  hypothetical
variations.  As  a matter  of fact, local  constraints on the  rate of
variation of $G$ can be derived, for example, from Lunar Laser ranging
\cite{Williams96,  Turyshev04},  whereas  constraints at  cosmological
distances  can be  derived,  amongst other  methods,  from the  Hubble
diagram of distant SNeIa \cite{Gaz02}.

On  the  other  hand,  in  the  framework of  models  with  a  varying
gravitational   constant  it  would   be  valuable   to  get   a  {\sl
phenomenological} expression for its variation.  That is, to obtain an
approximate form  of the function  $G(z)$, where $z$ is  the redshift.
Getting such description  is the goal of the  present paper.  We would
like to  note that fitting the  Hubble diagram of  SNeIa within models
with   a   variable    gravitational   constant   for   a   particular
parametrization and  $\Lambda =  0$ was studied  in \cite{Amendola99}.
Using  the  considerably   better  observational  data  available  now
\cite{Riess04} which  extend to much  larger distances we will  find a
more  accurate  approximation for  $G(z)$  for  a  larger interval  of
look--back times.  It is worth  mentioning as well that a procedure of
reconstruction  of a  general scalar--tensor  model (the  scalar field
potential and  the functional form of the  scalar-gravity coupling) of
dark  energy from  cosmological observational  data,  particularly the
luminosity distance, was first developed in \cite{Boisseau00}.

Once  the  phenomenological form  of  $G(z)$  is  obtained it  can  be
compared  with predictions  of cosmological  models  and/or contrasted
with  other astrophysical  observational  constraints.  For  instance,
models with  extra dimensions incorporate a natural  mechanism for the
space and  time variation of various fundamental  constants, which was
apparently studied for the first time in \cite{Forgacs79} and later on
in  a  number  of articles.   It  should  be  also mentioned  that  in
\cite{Lor03}  the relation  between  the time  variation  of the  fine
structure constant and that  of the gravitational constant was studied
for three classes  of theories --- namely, for  the pure Kaluza--Klein
theory,  for Einstein--Yang--Mills  theories and  for Randall--Sundrum
type models.  Using  the relation between $\alpha$ and  $G$ in a given
model  and the  phenomenological expression  for $G(z)$  one  can then
obtain an  estimate of the  variation of the fine  structure constant.
This   prediction  can   therefore   be  then   contrasted  with   the
observational  constraints on  the  variation of  $\alpha$, which  has
recently been a subject of intensive studies.  Particularly, using the
many  multiplet  method it  has  been  claimed \cite{Dzuba99,  Webb01,
Mur03} that  the fine structure  constant $\alpha$ was smaller  in the
past.  However,  a similar analysis carried out  in \cite{Chand04} and
in \cite{Srian04} using a different  line fitting code and data sample
of better quality shows that the measurements are consistent with zero
variation  within the  observational uncertainties  and, consequently,
these results do not support the claims by previous authors.

The plan of the paper  is the following. In Sect. \ref{sect:theory} we
outline a theoretical scheme for a variable gravitational ``constant''
and  derive a  generalization  of the  Hubble  law for  this case.   A
phenomenological  description of  the function  $G(z)$ which  fits the
Hubble diagram of SNeIa  is then found in Sect. \ref{sect:kinematics}.
In the  next section the  correlated variations of the  fine structure
constant  and of  the  gravitational constant  for  models with  extra
dimensions  is discussed.   Finally, in  Sect.  \ref{sect:conclusions}
our main conclusions are presented, followed by some discussion of our
most important results.

\section{Variation of $G$ in scalar-tensor theories}
\label{sect:theory}

Theories of gravity in which the gravitational ``constant'' $G$ varies
with time and cosmological models  based on them have been extensively
studied  in  the literature  (see,  for  example, \cite{Barrow97}  and
references  therein).  One  of   the  most  natural  and  relativistic
covariant ways to describe the variation of the gravitational constant
is  to  interpret it  as  a  scalar field  $\phi$.  This  can be  done
self--consistently  in  the framework  of  scalar--tensor theories  of
gravity  of the Jordan--Brans--Dicke  type \cite{BD61,  Jordan49} with
the action given by

\beq
S = \frac{1}{16 \pi} \int d^{4}x \sqrt{-g} 
\left( \phi {\cal R} + \frac{w(\phi)}{\phi} g^{\mu \nu} 
\partial_{\mu} \phi \partial_{\nu} \phi + 16 \pi L_{m} \right),   
\label{STT-action}
\eeq

\noindent  where  the  function  $w=w(\phi)$ determines  the  coupling
between the scalar field and gravity.

In considering the cosmic evolution  of the scale factor $a(t)$ of the
Friedmann--Robertson--Walker metric and of  the scalar field $\phi$ in
Eq.~(\ref{STT-action}) we assume for  simplicity that $w$ is constant.
Then, the Hubble  parameter $H \equiv \dot{a}/a$ is  determined by the
Friedmann equation \cite{Barrow97}

\beq
H^{2} \equiv \left( \frac{\dot{a}}{a} \right)^{2} 
=\frac{8\pi}{3 \phi} \rho - \frac{k}{a^2} - \frac{\dot{\phi}}{\phi} 
\frac{\dot{a}}{a} + \frac{w}{6} \frac{\dot{\phi}^{2}}{\phi^{2}} + 
\frac{\Lambda}{3},  \label{STT-H}
\eeq

\noindent where $\Lambda$ is the  cosmological constant and $k$ is the
curvature parameter.  We furthermore assume that the universe contains
a simple perfect fluid described by the equation of state

\beq
p =  (\gamma - 1)\rho.  \label{EOS} 
\eeq

\noindent   Eqs.~(\ref{STT-H})   and   (\ref{EOS})  and   the   energy
conservation condition  $\dot{\rho} + 3\gamma H  \rho = 0$  have to be
complemented    with   the    acceleration    equation   for    $\phi$
\cite{Barrow97}:

\beq
\ddot{\phi} + 3 H \dot{\phi} = \frac{8 \pi \rho}{2w+3} (4 - 3\gamma).  
\label{STT-phi}
\eeq

\noindent  In what  follows  we will  consider  $a$ and  $\phi$ to  be
functions of  the redshift  $z$.  To convert  time derivatives  to the
derivatives with respect to $z$ we use the standard relation:

\[
\frac{d}{dt}  =  -   H  (1  +  z)  \frac{d}{dz}.
\]

\noindent Denoting the $z$-derivatives  with prime we get relations of
the type $\dot{\phi} = - H (1+z) \phi'$.

By considering  the weak--field  limit in the  scalar--tensor theories
the  following relation  between  the gravitational  constant and  the
scalar field $\phi$ can be established \cite{BD61}

\beq
G(z) = \frac{4+2w}{3+2w} \frac{1}{\phi (z)}. 
\eeq

Using these expressions, the Hubble law --- given by Eq.~(\ref{STT-H})
--- can be written after some algebra in the following form

\beq
H^{2} = H^{2}_{0} g_{0} \frac{\hat{\Omega}_{M} 
\frac{G(z)}{G_{0}} (1+z)^{3\gamma} 
+ \hat{\Omega}_{R} (1+z)^{2} + \hat{\Omega}_{\Lambda}}{g(z)}. 
   \label{H-law}
\eeq

\noindent The function $g(z)$ is given by 

\beq
g(z) = 1 + (1+z) \frac{G'}{G} - \frac{w}{6} (1+z)^{2} 
\left( \frac{G'}{G} \right)^{2},     
\label{g-def}
\eeq

\noindent  where  $G_{0}=G(0)$  is   the  present--day  value  of  the
gravitational    constant,    $g_{0}=g(0)$,    and   the    parameters
$\hat{\Omega}_{M}$,  $\hat{\Omega}_{R}$  and  $\hat{\Omega}_{\Lambda}$
are related to the standard ratios

\[
\Omega_{M} \equiv \frac{8\pi G_{0} \rho_{0}}{3H_{0}^{2}}, \; \; \; 
\Omega_{R} \equiv -\frac{k}{a_{0}^{2} H_{0}^{2}}, \; \; \; 
\Omega_{\Lambda} \equiv \frac{\Lambda}{3 H_{0}^{2}}
\]

\noindent through the following relations 

\beq
\hat{\Omega}_{M} = \frac{\Omega_{M}}{g_{0}} \frac{3+2w}{4+2w}, \; \; \; 
\hat{\Omega}_{R} = \frac{\Omega_{R}}{g_{0}}, \; \; \; 
\hat{\Omega}_{\Lambda} = \frac{\Omega_{\Lambda}}{g_{0}}.
\eeq

\noindent From Eq.~(\ref{H-law}) it follows that 

\beq
\hat{\Omega}_{M} + \hat{\Omega}_{R} + \hat{\Omega}_{\Lambda} = 1. 
\label{Omega-rel}
\eeq

\noindent We would  like to note at this point  that a particular case
of    Eqs.~(\ref{H-law})--(\ref{Omega-rel})     was    discussed    in
\cite{Gazta01}.

Finally,  the  luminosity  distance  $d_{L}$  is  calculated  via  the
standard formula, which in the flat case has the form

\beq
d_{L}= c (1+z) \int_{0}^{z} \frac{du}{H(u)}.    \label{dL}
\eeq

\noindent Similar to the way in which it was done in \cite{Gaz02}, for
the  calculation  of the  Hubble  diagram of  SNeIa  we  will use  the
Chandrasekhar  mass model  for the  SNeIa light  curve.   According to
this,  the  peak  luminosities   of  SNeIa  are  proportional  to  the
Chandrasekhar mass  ($L \propto M_{\rm Ch}$) and,  therefore, scale as
$L  \propto G^{-3/2}$.   This result  has been  validated  by detailed
numerical  calculations of exploding  white dwarfs.   As a  result the
apparent magnitude is given by

\beq
m(z) = M_{0} + 5 \log \frac{d_{L}(z) H_{0}}{c} + 25 + 
\frac{15}{4} \log \frac{G(z)}{G_{0}},    \label{m-rel}
\eeq

\noindent  where  $M_{0}$  is  the  absolute magnitude.  We  will  use
Eqs.~(\ref{dL}) and (\ref{m-rel}) to fit the Hubble diagram of distant
SNeIa with a certain parametric representation for $G(z)$.

\section{A phenomenological description of the variation of the
         gravitational constant}
\label{sect:kinematics}

As it has been already commented before, our goal is to obtain an {\sl
empirical} description of the  variation of the gravitational constant
as inferred from  the observational data of SNeIa,  including the most
recent and reliable datasets  \cite{Riess04}.  For this purpose we use
a simple  phenomenological expansion of the function  $G(z)$ in powers
of $z$:

\beq
G(z) = G_{0} (1 + p_{1} z + p_{2} z^{2} + p_{3} z^{3} + 
{\cal O}(z^{4})).      
\label{G-exp}
\eeq

\noindent  Such approximation  is  in the  spirit of  phenomenological
descriptions of  the scale factor, deceleration  parameter or equation
of state emloyed previously  by other authors \cite{Riess04, Turner02,
Visser04}.

The  first coefficient  in Eq.~(\ref{G-exp}),  $p_{1}$,  is determined
from experimental  bounds on the time derivative  of the gravitational
constant    at    the    present    time    $(\dot{G}/G)_{0}    \equiv
(\dot{G}/G)_{t=t_{\rm  now}}$.   For  convenience we  translate  these
bounds in the ones on $\eta \equiv G'(0)/G_{0}$ using the relation

\[
    \frac{G'(0)}{G_{0}} = - \frac{1}{H_{0}} 
  \left( \frac{\dot{G}}{G} \right)_{t=t_{\rm now}}.
\]

\noindent  For this estimate  we will  take the  present value  of the
Hubble constant to be $H_{0}=63$~km/s/Mpc.

There are  a number of constraints on  $(\dot{G}/G)_{0}$ obtained from
very  different  observations  and  methods  ---  see,  for  instance,
\cite{Uzan03} and \cite{GIKL04}  and references therein.  For example,
the Lunar Laser ranging  experiments yield $|\dot{G}/G|_{0} < 8 \times
10^{-12} \;\mbox{yr}^{-1}$  \cite{Williams96, Turyshev04}, whereas the
improved constraints  on the post--Newtonian parameters  give an upper
bound  of $10^{-14} \;\mbox{yr}^{-1}$  \cite{Bertotti03}. Summarizing,
one  can see  that the  parameter  $\eta$ can  take values  satisfying
roughly $|\eta| \leq 0.01$, which  is similar to the estimate obtained
in \cite{Amendola99}.  Actually, as far as $\eta$ is small enough, the
values of  the other  coefficients depend very  weakly on  its precise
value.   Moreover,  we have  checked  that  our  final result  is  not
sensitive to the  value of $\eta$ within the interval  $-0.01 < \eta <
0.01$.  Since $\eta = p_{1}$ the  bound on the local rate of variation
of  $G$ determines  the  linear term  in  Eq.~(\ref{G-exp}).  For  the
forthcoming  analysis  we  adopt $p_{1}=\eta=-0.01$.   

In what  follows we truncate the expansion  given in Eq.~(\ref{G-exp})
at  a  certain  order,  substitute  this  polynomial  expression  into
Eq.~(\ref{H-law}), calculate the apparent magnitudes of SNeIa as given
by Eq.~(\ref{m-rel}) in terms  of the coefficients $p_{i}$ and compare
them with the Hubble diagram based on the observational data of SNeIa.

Let us  consider the  case of the  flat and matter  dominated universe
without  the  cosmological  constant.  Namely, we  set  $\Omega_{R}  =
\Omega_{\Lambda}   =   0$   and   $\gamma=1$  in   the   Hubble   law,
Eq.  (\ref{H-law}).  An  important   observation  is  that  with  such
assumptions the value of the  ratio $\Omega_{M}$ turns out to be fixed
by the value of $p_{1}$. Indeed, from (\ref{g-def}) one gets

\beq
\Omega_{M}= \frac{4+2w}{3+2w} g_{0} = \frac{4+2w}{3+2w} 
\left( 1 + p_{1} - \frac{w}{6} p_{1}^{2} \right). \label{Omega-M}
\eeq 

The value  of the Brans--Dicke  parameter $w$ depends on  the specific
model.     For   example,    in   the    case    of   multidimensional
Einstein--Yang--Mills models  with $d$ extra  dimensions, discussed in
the next section, $w=(d-1)/d$ \cite{Kub89}.  Having in mind this class
of  models we  take $w  \sim 0.5  \div 1$.   The exact  value  of this
parameter does  not affect our final  result in a  significant way. In
the   case  of   models   obtained  by   dimensional  reduction   from
multidimensional  theories with  six  extra dimensions,  which can  be
motivated  by the  string  theory, from  Eq.~(\ref{Omega-M}) one  gets
$\Omega_{M} \approx 1.2$.

\begin{figure}[t]
\centering
\includegraphics[clip, width=11 cm]{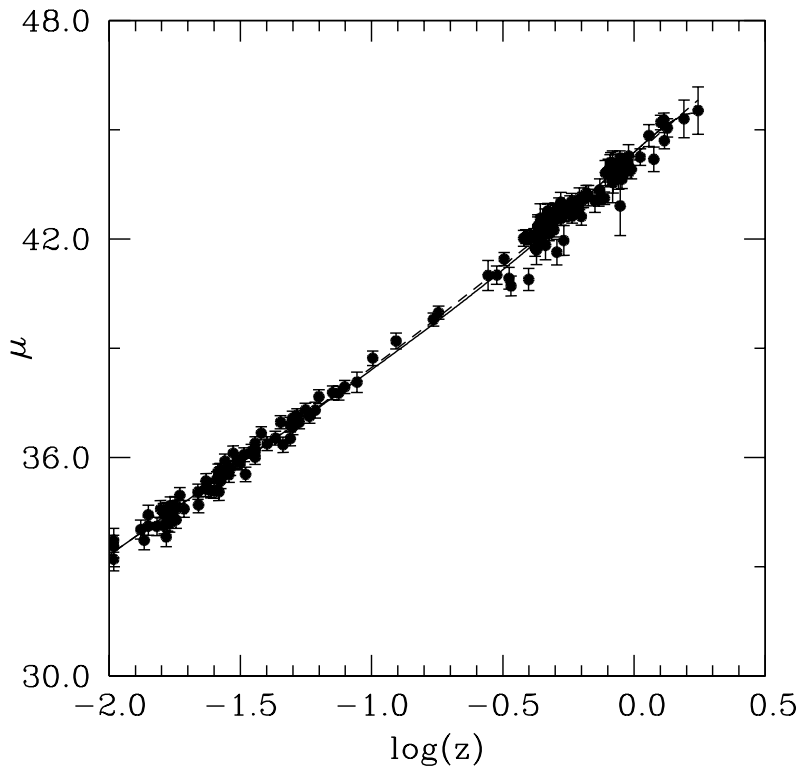}
\caption{The observational Hubble  diagram of distant supernovae (dots
      with their corresponding  error bars), the best fit  curve to it
      in the model with  the variable gravitational constant $G(z)$ as
      given by Eq.~(\ref{G-exp}) (solid line).}
\label{fig:hubble}
\end{figure}

Fitting  the Hubble  law to  the  supernova dataset  with a  quadratic
polynomial for $G(z)$ --- i.e.,  with only one free parameter, $p_{2}$
--- does   not   give  satisfactory   results.    To   get  a   better
phenomenological  approximation we  consider a  cubic polynomial  as a
parametrization of the  function $G(z)$ in Eq.~(\ref{G-exp}).  Varying
$p_{2}$  and $p_{3}$  we  obtained that  the  best fit  to the  Hubble
diagram of SNeIa  is achieved when the values  of the parameters $p_2$
and  $p_3$   in  Eq.~(\ref{G-exp})   are  $p_{2}  \approx   0.34$  and
$p_{3}=-0.17$,   respectively. 

Fig.~\ref{fig:hubble}  shows  the   observational  Hubble  diagram  of
distance moduli for  SNeIa based on the data  of \cite{Riess04} (their
gold  sample). Overplotted  is the  best fit  curve for  the predicted
distance modulus

\beq
\mu_{\rm th}(z) \equiv 5 \log d_{L}(z) + 25 + 
\frac{15}{4} \log \frac{G(z)}{G_{0}},   \label{mu-th}  
\eeq

\noindent  calculated in  our  model with  the variable  gravitational
constant $G(z)$, Eq. (\ref{G-exp}). For these calculations we used the
present Hubble parameter $H_{0} = 63\; \mbox{km/s/Mpc}$.  As it can be
seen from Fig.~\ref{fig:hubble}  the agreement between the theoretical
fit and the observations is quite good.

The likelihood  for the parameters  $p_{2}$ and $p_{3}$  is determined
from the $\chi^{2}$-analysis with

\beq
\chi^{2}(p_{2},p_{3})= \sum_{i} 
\frac{\left[ \mu_{\rm th}(z_{i}) - \mu_{\rm obs}^i
\right]^{2}}{{\sigma_{\mu}^i}^{2}},                 
\label{chi-2}
\eeq

\noindent  where $\mu_{\rm  th}(z)$  is given  by Eq.   (\ref{mu-th}),
$\mu_{\rm obs}^i$  are the observational data for  the distance moduli
and $\sigma_{\mu}^i$ are the  uncertainties in the individual distance
moduli. Definition (\ref{chi-2}) of $\chi^{2}$ is analogous to the one
in Ref. \cite{Riess04}.  We would like  to add that the best fit cubic
polynomial $G(z)$ yields $\chi^{2}=199$,  for comparision the value of
$\chi^{2}$  obtained  for  the  cosmic  concordance  model,  which  is
$\chi^{2}=178$.

\begin{figure}[t]
\centering
\includegraphics[clip, width=11 cm]{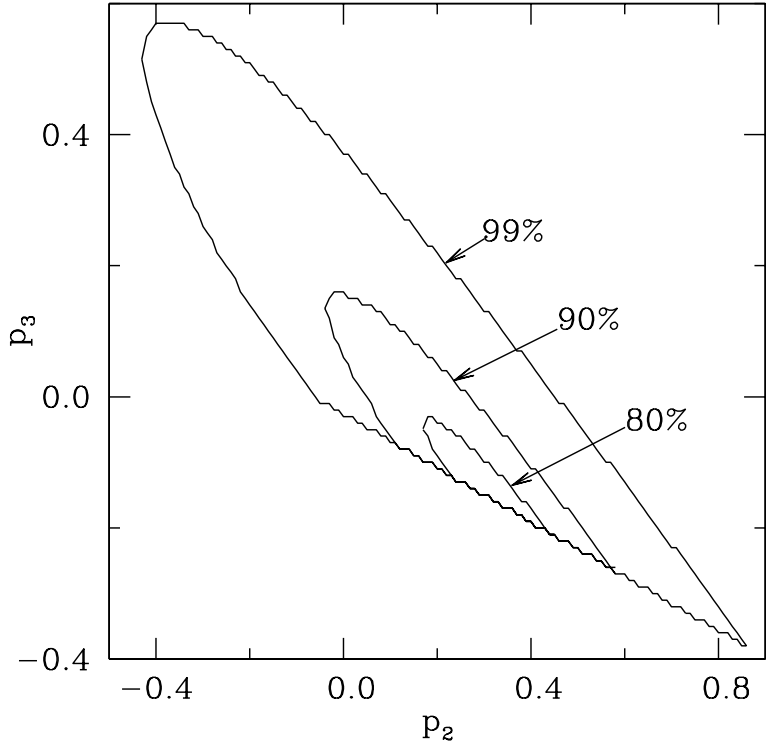}
\caption{Joint  confidence intervals  for the  coefficients  $p_2$ and
         $p_3$  of   the  polynomial  fit   to  $G(z)$  as   given  in
         Eq.~(\ref{G-exp}).}
\label{fig:contour}
\end{figure}

Fig.~\ref{fig:contour}  shows the joint  confidence intervals  for the
fit  to the SNeIa  observational data.  The analysis  was done  in the
region of the parameters  $p_{2},p_{3}$ such that $H^{2}(z)>0$ for all
redshifts in the  interval $0 < z < 1.7$. The  bottom boundary of this
region  is seen  in Fig.~\ref{fig:contour}.  To be  more  precise, for
values  of these  parameters below  the boundary  the  function $g(z)$
defined  by Eq. (\ref{g-def})  becomes negative  as $z$  approaches to
$z=1.7$. The  best fit values of  $p_{2}$, $p_{3}$ are  in fact rather
close to this boundary.

\begin{figure}[t]
\centering
\includegraphics[clip, width=11 cm]{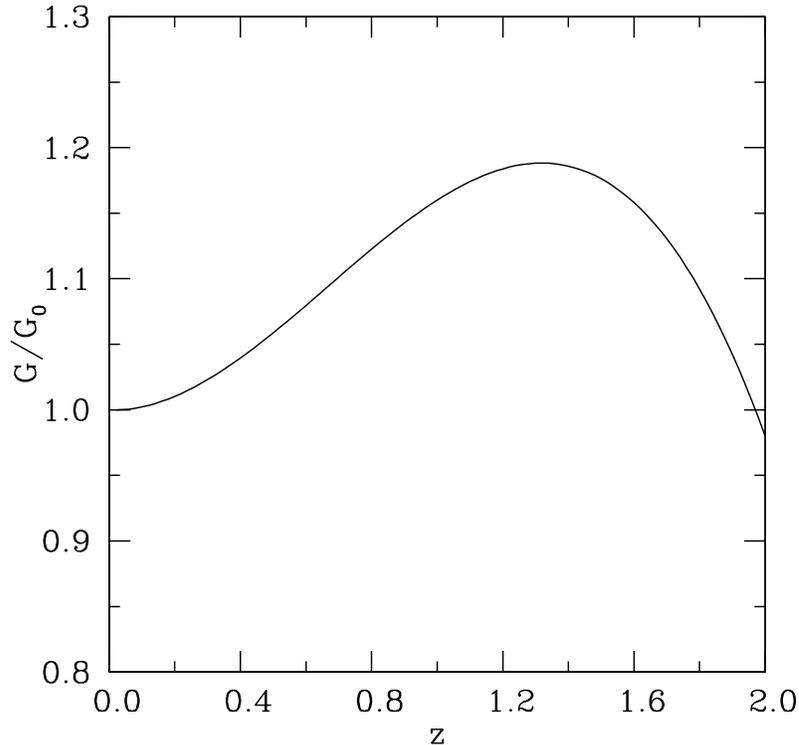}
\caption{The  function  $G(z)$  represented  by  cubic  polynomial  of
      Eq.~(\ref{G-exp}) with the coefficients determined from the best
      fit to the SNeIa Hubble diagram.}
\label{fig:G}
\end{figure}
 
The  plot of  the cubic  polynomial for  $G(z)$ with  the coefficients
found  above  is  shown  in  Fig.~\ref{fig:G}.   The  phenomenological
expression for $G(z)$ based on  the SNeIa data suggests that the value
of  the gravitational constant  was larger  in the  past.  To  be more
precise, $G(z)  > G_{0}$ in the  interval from $z \approx  0.03$ to $z
\approx  1.97$. It can  be seen  from Fig.~\ref{fig:G}  that for  $z >
0.03$ as one moves towards  larger redshifts the function $G(z)$ first
grows, reaches  its maximum $G_{\rm  max}=1.19\, G_{0}$ at $z  = 1.32$
and  then   steadily  decreases.   Of   course,  the  phenomenological
expression  for $G(z)$  obtained  here is  approximate and  presumably
makes sense  as far  as the last,  cubic term in  Eq.~(\ref{G-exp}) is
smaller then the previous quadratic  term, that is roughly for $z \leq
2$.

The form obtained  here for the function $G(z)$  should be constrasted
with  existing  constraints  and/or   compared  with  results  on  the
variation of $G$ in  various cosmological models.  As an illustration,
in   the  following   section   $G(z)$  will   be   considered  as   a
phenomenological input  in a  multidimensional model and  a prediction
for the variation  of the fine structure constant  will be derived and
analyzed.

\section{Variation of $G$ and $\alpha$ in multidimensional models}
\label{sect:KK}

The  time variation of  the fundamental  constants within  models with
extra  dimensions   has  been  considered   in  a  number   of  papers
\cite{Forgacs79}, \cite{var-Xdim}--\cite{KOP},  just to mention  a few
of them.   Additionally, in \cite{Lor03} the  correlated variations of
the  fine structure constant  and of  the gravitational  constant were
analyzed.  In this paper it was shown that in the framework of certain
multidimensional  models there  exists a  robust relation  between the
time derivatives  of $\alpha$ and  $G$ which quite generically  can be
written as

\[
\frac{\dot{\alpha}}{\alpha} = \beta \frac{\dot{G}}{G}, 
\]

\noindent where the factor $\beta$  is model dependent and, in general, 
may depend on the   scale  (or  size)  $R$  of   the  space  of  extra
dimensions. For the case of constant $\beta$ a similar relation between 
the derivatives with respect to the redshift holds:

\beq
\frac{\alpha'}{\alpha} = \beta \frac{G'}{G}. 
\eeq

\noindent Integrating this equation one gets  

\beq
\frac{\Delta \alpha}{\alpha}(z) \equiv \frac{\alpha (z) - \alpha_{0}}
{\alpha_{0}} = 
\left( \frac{G(z)}{G_{0}} \right)^{\beta}-1,   
\label{alpha-G}
\eeq

\noindent where  $\alpha_{0}=\alpha (0)$ is  the present day  value of
the fine structure constant.

To  be  specific,  let   us  consider  the  case  of  multidimensional
Einstein--Yang--Mills  theories. In  this case  it can  be  shown that
$\beta=1$.  Using the cubic  polynomial of Eq.~(\ref{G-exp}), with the
coefficients  determined   in  the  previous   section,  the  redshift
dependence of $\alpha$  can be obtained and, from it, the behaviour of
$\Delta \alpha/\alpha$  predicted for such theory can  be derived.  In
particular,  it can  be seen  that at  $z=0.5$ this  ratio  is $\Delta
\alpha/\alpha \approx  0.06$.  This theoretical prediction  is at odds
with  the known  constraints  on the  rate  of variation  of the  fine
structure  constant.   More specifically,  the  latest  analysis of  a
Keck/Hires sample of quasar  absorption lines using the many multiplet
method gives

\beq 
\frac{\Delta \alpha}{\alpha} = (-0.54 \pm 0.12) \times 10^{-5} 
\label{delta-a}
\eeq 

\noindent for  $z$ in  the range  $0.5 < z  < 3$  \cite{Mur03}.  Note,
moreover,  that  these authors  obtained  that  the  value of  $\Delta
\alpha$ is {\sl negative}, whereas the results derived from the Hubble
diagram  of SNeIa  predict $\Delta  \alpha>0$  for the  same range  of
redshifts in the cosmological theories under consideration.  Moreover,
as already mentioned in Sect.  1, the observational results leading to
a  non--zero  rate  of  variation  of $\alpha$  have  been  challenged
recently \cite{Chand04, Srian04}.   These studies provide much tighter
constraints on the  rate of variation of the  fine structure constant.
In particular,  the constraint  $\Delta \alpha /  \alpha =  (-0.06 \pm
0.06) \times  10^{-5}$ was  obtained,  which is  consistent with  zero
variation.  In summary, neither the sign of our theoretical prediction
nor its order of magnitude coincide with the above bounds.  In fact, a
similar conclusion was formulated in \cite{Lor03}.

The  constraints  studied  before  correspond  to  a  redshift  $z\sim
0.5$.  The best local  ($z\sim 0.0$)  bound on  the time  variation of
$\alpha$   is   that   obtained   from  the   Oklo   natural   nuclear
reactor:

\[
\frac{\Delta \alpha}{\alpha} = (0.15 \pm 1.05) \times 10^{-7} 
\] 

\noindent   at   $z   \approx   0.15$   \cite{Oklo}.    The   obtained
phenomenological  formula  yields  $\Delta  \alpha  /  \alpha  \approx
0.006$. Actually, as one can see from Eq.~(\ref{alpha-G}), for $\Delta
\alpha / \alpha$ to fit the Oklo constraint at $z = 0.15$ the value of
the  parameter  $\beta$  should  be  of the  order  of  $|\beta|  \sim
10^{-5}$,  which  is  the  same  to  say that  $\alpha  (z)$  must  be
practically independent of $G(z)$.

\section{Conclusions and discussion}
\label{sect:conclusions}

We have  studied the possibility  of fitting the  observational Hubble
diagram  of SNeIa  assuming  cosmological models  of  a flat  universe
without  cosmological  constant   but  with  a  varying  gravitational
``constant'', $G(z)$.  The function  $G(z)$ was represented by a cubic
polynomial,  Eq.~(\ref{G-exp}), parametrized with  three coefficients,
$p_{1}$, $p_{2}$  and $p_{3}$. The linear order  coefficient was fixed
by the  constraints on the present  day rate of the  time variation of
$G$.  The other two coefficients  were determined from the best fit to
the  Hubble  diagram.   Finally  we  arrived  at  the  following  {\sl
phenomenological} expression:

\beq
G(z) = G_{0} \left(1 - 0.01 z + 0.34 z^{2} - 0.17 z^{3} \right),      
\label{G-final}
\eeq

\noindent  where  $G_{0}=G(0)$  is   the  present  day  value  of  the
gravitational  constant. It  is important  to mention  that  the sharp
boundary  at   the  bottom  of  the  joint   confidence  intervals  in
Fig.~\ref{fig:contour}  is due  to  the restriction  that no  negative
values  of  $H^{2}(z)$  occur  in  the  range of  redshifts  $0<  z  <
1.7$.  Values of  $p_3$  smaller  that those  of  the bottom  boundary
produce $g(z) <  0$ (see Eq. (\ref{H-law})) for some  values of $z$ in
this interval and are consequently discarded.

In  this  paper  we  limited  ourselves  to  a  cubic  polynomial  for
$G(z)$. Such  choice of approximation  is motivated by the  results on
two--parametric descriptions of astrophysical characteristics obtained
from  the   same  datasets  in   \cite{Riess04},  \cite{Turner02}  and
\cite{Visser04}, which suggest that  the available data does not allow
a good determination of higher order parameters.  The phenomenological
determination of $G(z)$ from SNeIa data suggests that the value of the
gravitational constant was higher than the present day one for $0.03 <
z < 1.97$ with the maximal value $G_{\rm max}=1.19\, G_{0}$ reached at
$z = 1.32$.  This change  from the growing to the decreasing behaviour
of $G(z)$  is a manifestation  of the change  in the behaviour  of the
observational data  with the redshift. The latter  feature was studied
in Ref. \cite{Riess04} in terms of the deceleration $q(z)$ represented
by a  linear polynomial with two parameters.   Using the observational
data for distant SNeIa  the change of sign of $q(z)$ at  $z = 0.46 \pm
0.13$ was  discovered.  This  is interpreted as  an indication  of the
change from the epoch of acceleration of the evolution of the Universe
to the epoch of deceleration as $z$ increases.

The  features  of  the   approximation  $G(z)$  obtained  here  should
definitely be compared, within the  domain of its validity, with other
cosmological   and  astrophysical   bounds   and  restrictions.    The
discrepancy between  the theoretical  prediction for the  variation of
the  fine  structure  constant  obtained  in  the  models  with  extra
dimensions and  the existing observational  constraints, provided that
the  latter  are  solid  and  confirmed,  indicates  that  either  the
multidimensional   models  considered   here   are  phenomenologically
unsatisfactory or the very hypothesis of the variability of $G$ is not
correct.

\section*{Acknowledgements} 
This  work has  been partially  supported  by the  Spanish MEC  grants
AYA2005--08013--C03--C01 and  C02,  by the AGAUR  and by  the European
Union  FEDER  funds.   The  work  of  Y.~K.   was   supported  by  the
UR.02.03.028  grant of  the Programme  ``Universities of  Russia'' and
grant 04--02--16476 of the Russian Fund for Basic Research.


\end{document}